\documentclass[twocolumn,nobibnotes,showpacs,preprintnumbers,amsmath,amssymb]{revtex4}
\pdfoutput=1
\usepackage{graphicx}
\usepackage{dcolumn}
\usepackage{bm}

\newcommand{\be}{\begin{eqnarray}}
\newcommand{\ee}{\end{eqnarray}}
\newcommand{\bea}{\begin{eqnarray}}
\newcommand{\eea}{\end{eqnarray}}

\newcommand{\meV}{{~\rm meV}}

\newcommand{\MeV}{{~\rm MeV}}

\newcommand{\mphi}{m_\phi}

\begin{document}

\title{Muonic hydrogen and MeV forces}
\author{David Tucker-Smith$^{(a,b,c)}$, and Itay Yavin$^{(b)}$}

\affiliation{(a) Department of Physics, Williams College, Williamstown, MA 01267, USA \\ (b) Center for Cosmology and Particle Physics, Department of Physics, New York University, New York, NY 10003\\ (c) School of Natural Sciences, Institute for Advanced Study, Princeton, NJ 08540}


\begin{abstract}
We explore the possibility that a new interaction between muons and protons is responsible for the discrepancy between the CODATA value of the proton radius and the value deduced from the measurement of  the Lamb shift in muonic hydrogen.  We show that a new force carrier with roughly MeV mass can account for the observed energy shift as well as the discrepancy in the muon anomalous magnetic moment.  However,  measurements in other systems constrain the couplings to electrons and neutrons to be suppressed relative to the couplings to muons and protons, which seems challenging from a theoretical point of view. One can nevertheless make predictions for energy shifts in muonic deuterium, muonic helium, and true muonium under the assumption that the new particle couples dominantly to muons and protons. 
\end{abstract}
%
%

\maketitle

\section{Introduction}

A recent publication~\cite{Pohl:2010zz} announced a measurement of the Lamb shift  in muonic hydrogen that seems to require a value of the proton's radius, $r_p=0.84184(67) $ fm,  which differs by five standard deviations from the value given in the CODATA compilation~\cite{Mohr:2008fa}, $r_p=0.8768(69)$ fm.  
It is of course possible that the  reason for the different values of $r_p$ extracted from muonic and electronic hydrogen lies within the standard model, involving subtle QED and/or hadronic effects.  It is nevertheless worth considering whether new physics could be the explanation.  Here we explore the possibility that the discrepancy arises from the existence of a roughly MeV-mass force carrier that couples the muon to the proton. As we will see, such a force can also resolve the long-standing discrepancy between theory and observation  in $(g-2)_\mu$ measurements. However, atomic precision measurements and neutron scattering experiments at low energies  constrain the coupling of this new force to electrons and neutrons, respectively. It is not difficult to construct models that effectively decouple the force carrier from the electron (by arranging for the particle to couple to mass) {\em or} the neutron (by arranging for the particle to couple to charge), but decoupling it from both is  a  more serious challenge to model building attempts. Taking the force to act on muons and protons only,   it is still possible to make predictions for energy shifts in related systems -- muonic deuterium, muonic helium, and true muonium -- with minimal model-dependence.  

\section{Contributions to the Lamb Shift in Hydrogenic Systems} 

Independent of whether the new force is mediated by a scalar or a vector-boson, the non-relativistic potential between the proton and muon can be written as 
\be
V_\phi(r) = (-)^{s+1}  \left(\frac{g_\mu g_p}{e^2} \right) \frac{\alpha e^{-\mphi r}}{r},
\ee
where $\mphi$ is the mass of the force carrier, $s$ is its spin, and $g_\mu$ and $g_p$ are its couplings to the muon and proton, respectively. A similar expression holds for the electronic-hydrogen system.  Note, however, that the sign of the potential may be different depending on the relative sign between $g_e$ and $g_\mu$. 

This potential gives an additional contribution to the Lamb shift in the 2S$_{1/2}$-2P$_{3/2}$ transition, which using first-order perturbation theory is given by
\be
\label{eqn:energyshift}
\delta E_\phi &=&  \int dr r^2 V_\phi(r) \left(|R_{20}(r)|^2 - |R_{21}(r)|^2 \right) \\\nonumber &=& (-)^{s+1}\frac{\alpha}{2a_\mu^3} ~\left(\frac{g_\mu g_p}{e^2} \right)\frac{f(a_\mu \mphi)}{\mphi^2},
\ee
with $f(x) = x^4/(1+x)^4$. Here $a_\mu = (\alpha m_{\mu p})^{-1}$ is the Bohr radius of the system to leading order and $m_{\mu p}$ is the reduced mass of the $\mu$-$p$ system. A similar expression holds for the $e$-$p$ system, but the Bohr radius is a factor $\approx m_\mu/m_e$ larger. This expression is convenient for a direct comparison with the leading order contribution from the proton radius, 
\be
\delta E_p = \frac{2\alpha}{3n^3 a_\mu^3}\langle r_p^2 \rangle, 
\ee
where $r_p$ is the proton's radius, and $n(=2)$ is the principle quantum number.

If everything else is the same between the $e$-$p$ and $\mu$-$p$ systems, then since $f(x)$ is a monotonically increasing function that asymptotes to unity at large values of $x$, the resulting energy shift in the  $e$-$p$ system is always \textit{larger} than the corresponding shift in the $\mu$-$p$ system \cite{Jaeckel:2010xx}.  If the force is attractive in both systems, the apparent proton radius will always appear \textit{smaller} in the $e$-$p$ system, contrary to observations. 
An attractive force must therefore couple more strongly to muons than to electrons if it is to explain the discrepancy in the proton-radius determination.  

A different possibility is that $g_e$ is not suppressed relative to $g_\mu$, but the force is repulsive, leading to  a larger apparent proton-radius in the $e$-$p$ system. This possibility is consistent with the $(g-2)_e$ constraint discussed in the next section. However, since the effects of such a light force are suppressed at higher momentum transfer, this possibility seems in tension with the  value of the proton radius extracted from scattering data which generally also imply a larger proton radius~\cite{Hill:2010yb}. Therefore, for the purpose of this paper, we concentrate on the possibility that it is a new attractive force that modifies the $\mu$-$p$ system. 

\section{Contributions to $(g-2)_{e,\mu}$} 

The scalar and  vector-boson contributions to the electron and muon anomalous magnetic moments are~\cite{Schwinger:1948iu, Jackiw:1972jz}, 
\be
\Delta a_l = \frac{\alpha}{2\pi}\left(\frac{g_\mu}{e}\right)^2 \xi\left(\mphi/m_l\right),
\ee
where $m_l$ is the mass of the electron or muon, and
\be
\xi(x)_{\rm scalar} &=& \int_0^1 \frac{(1-z)^2(1+z)}{(1-z)^2+ x^2 z} dz\\
\xi(x)_{\rm vector} &=& \int_0^1 \frac{2 z(1-z)^2}{(1-z)^2 + x^2 z} dz. 
\ee
For $m_l \gg \mphi$ we have the asymptotic behaviors $\xi_{scalar} \rightarrow 3/2$ and $\xi_{vector} \rightarrow 1$. 

We begin with  the electron system. As emphasized in ref.~\cite{Pospelov:2008zw}, the $(g-2)_e$ measurement is currently used to define the fine-structure constant $\alpha$. The additional contribution to $(g-2)_e$ therefore acts as a shift of the fine-structure constant as $\Delta \alpha = 2\pi \Delta a_e$. Comparing this correction to measurements made in Rb and Cs atoms~\cite{Clade:2006zz,Gerginov:2006zz}, the shift in $\alpha$ must not exceed 15~ppb, which constrains the coupling to electrons as~\cite{Pospelov:2008zw}
\be
\left(\frac{g_e}{e}\right)^2 \xi\left(\mphi/m_l\right) < 15\times 10^{-9}.
\label{eqn:g-2_e}
\ee
For $\mphi\approx\MeV$ this constraint translates to $g_e/e \lesssim 2.3 \times10^{-4}$ and $g_e/e \lesssim 4.0 \times10^{-4}$ for scalar and vector mediators, respectively. The constraint is weakened for larger values of $\mphi$. 

\begin{figure}
\begin{center}
\includegraphics[width=0.45 \textwidth]{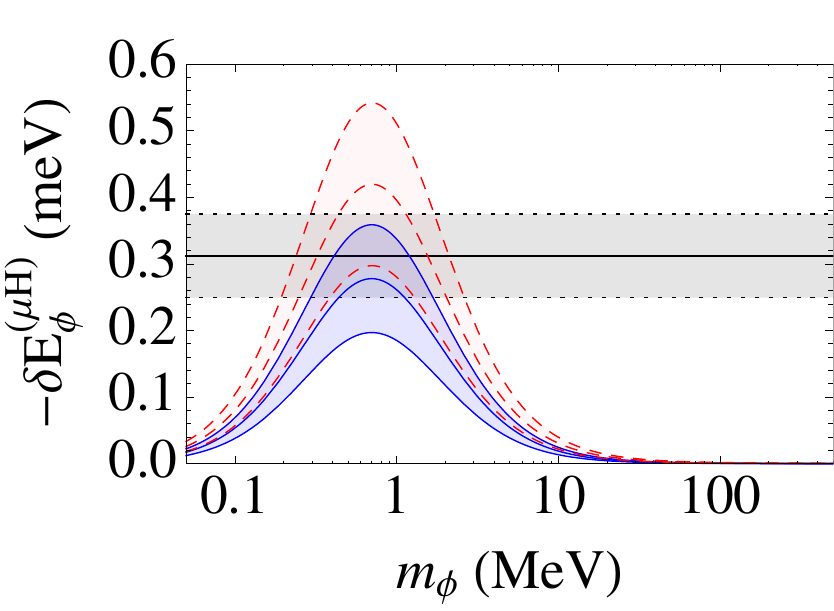}
\end{center}
\caption{The contribution to the energy shift in muonic hydrogen, Eq.~(\ref{eqn:energyshift}), plotted against the mass of the mediator. In the central solid-blue curve we require the coupling to the muon, $g_\mu$, to be such that the scalar contribution to $(g-2)_\mu$ equals the  theoretical deficit. In the upper/lower solid-blue curve  the scalar contribution to $(g-2)_\mu$ is determined to be  $\pm1$  s.d. away from the theoretical deficit. The vector case is similarly given by the dashed-red curves. The coupling to the proton is fixed at $g_p = g_\mu$. The solid horizontal line is the discrepancy between the experimentally measured value of the energy and the theoretical prediction assuming the CODATA value~\cite{Mohr:2008fa} for the proton-radius, $r_p = 0.8768~\rm{fm}$. The dotted horizontal lines represent the $\pm1$ s.d. uncertainty about this value.}
\label{fig:deltaEvsM}
\end{figure}

More important for the purpose of a direct comparison with the Lamb shift in the $\mu$-$p$ system is the constraint coming from measurement of $(g-2)_\mu$~\cite{Bennett:2006fi}. At present, the theoretical prediction for $a_\mu^{th}$ seems to indicate a deficit of $302(88)\times10^{-11}$ compared with the experimental value $a_\mu^{exp}$~\cite{Passera:2008hj}. 
For a given mass $\mphi$ we can extract the values of $g_\mu$ that bring the theoretical and experimental values of  $(g-2)_\mu$ into agreement.  
In fact,  these values are insensitive to $\mphi$ provided $\mphi \ll m_\mu$ is satisfied, giving $g_\mu/e\approx1.6\times 10^{-3}$ for a vector and $g_\mu/e\approx1.3\times 10^{-3}$ for a scalar. 
In Fig.~\ref{fig:deltaEvsM} we plot the 2S$_{1/2}$-2P$_{3/2}$ energy shift, Eq.~(\ref{eqn:energyshift}), against the mass of the mediator, $\mphi$, fixing the coupling to muons in this way. For the purpose of the plot we take $g_p = g_\mu$, but the result for other choices is easily obtained, as the energy shift is proportional to $g_p$.

As the plot indicates, a new force with a mass $\approx\MeV$ and a coupling to muons that explains the discrepancy in the muon anomalous magnetic moment can also give the required energy shift in muonic hydrogen to reconcile the proton-radius extracted from this system with the one extracted from hydrogen and electron-proton scattering. The mediator mass that gives the maximum energy shift is  near an $\MeV$ because it is essentially determined by the Bohr radius of the $\mu$-$p$ system, $a_\mu^{-1} = 0.69\MeV$. The choice $g_p\approx g_\mu$ favors $\mphi\approx\MeV$, but  a different mass can be accommodated by increasing $g_p$ accordingly. For $\mphi\gg \MeV$ ($\mphi\ll \MeV$) the required coupling becomes large and scales as $\mphi^2$ ($\mphi^{-2}$).

For $\mphi \approx \MeV$ the coupling to muons necessary to explain both discrepancies turns out to be close to the muon mass divided by the electroweak scale, $m_\mu/v =  4.3\times 10^{-4}$. A Higgs-like coupling  proportional to the mass would resolve any tension with the constraint from $(g-2)_e$ on the electron coupling,  Eq.~(\ref{eqn:g-2_e}). However, it implies that the coupling to neutrons is comparable to the coupling to protons, $g_n\approx g_p$, which seems severely constrained by neutron scattering experiments, as we discuss in the following section. 

\section{Constraints}

The most model-independent constraint on an MeV-scale interaction between muons and protons comes from measurements of the 3d$_{5/2}$-2p$_{3/2}$ transitions in $^{24}$Mg and $^{28}$Si \cite{Beltrami:1985dc}. A weighted-average between the results from $^{24}$Mg and $^{28}$Si was used to obtain the limit, $\left(\lambda_{exp}-\lambda_{QED}\right)/\lambda_{QED} = (-0.2\pm3.1)\times 10^{-6}$. For $\mphi\approx\MeV$ this translates to a $95\%$ CL limit  $g_p g_\mu/e^2 < 3.1\times 10^{-6}$, assuming coupling only to protons. This measurement therefore allows the couplings necessary to explain the muonic hydrogen and $(g-2)_\mu$ discrepancies. Mild tension does arise if one assumes that  the new force carrier also couples to neutrons since the bound improves by a factor of 2. However, as we now discuss, such a coupling to the neutron is much better constrained by neutron scattering experiments. 

A new force carrier with an MeV mass that couples to neutrons produces sizable corrections to the scattering cross-section of neutrons on heavy nuclei. This was first realized by Barbieri and Ericson~\cite{Barbieri:1975xy} who used the results of an old experiment~\cite{Aleksandrov:1966} on the polarizability of the neutron to set a limit on any additional force carrier that interferes with the strong interaction amplitude. They showed that such a force will contribute to the angular dependence of the differential cross-section in a distinct manner as compared with the contribution from the strong interactions. Assuming that the new force couples equally to both protons and neutrons, the bound on the nucleon coupling is
\be
g_n \lesssim 2\times 10^{-5} \left(\mphi/\MeV\right)^2.
\label{eqn:neutronbound}
\ee
This is about an order of magnitude smaller than the necessary coupling to muons and protons discussed above. A more recent analysis arrived at a similar result~\cite{Schmiedmayer:1991zz,Leeb:1992qf} although the claimed precision of that experiment  was later questioned (see Ref.~\cite{Wissmann:1998ta} and references therein). 

The bound in Eq.~(\ref{eqn:neutronbound}) arises from an interference term between the strong amplitude and the amplitude due to the new force. As such, it may be susceptible to cancellations involving other parts of the amplitude, and it also depends on  the relative phase of the strong and new-physics amplitudes. 
Nevertheless, the constraint is sufficiently strong to disfavor  $g_n\approx g_p$. The bound on the neutron coupling would have to be invalidated by more than an order of magnitude to allow for a simultaneous explanation of $(g-2)_\mu$ and the muonic hydrogen results while requiring $g_n\approx g_p$, a possibility which we therefore eschew in this letter.   

There are several other constraints on such a light boson, but they all involve further assumptions about its couplings to matter. Refs.~\cite{Kohler:1974zz,Freedman:1984sd} use $^{16}$O and $^4$He atoms to search for $0^+\rightarrow 0^++\phi$ transitions to constrain light bosons with Higgs-like couplings. These bounds are not very useful for $\mphi \approx \MeV$ and in any case depend sensitively on the decay properties of $\phi$.  For example, the bounds do not apply  if $\phi$ decays promptly, or if $\phi$ is too light to decay into an electron-positron pair. Ref.~\cite{Davier:1989wz} has searched for a light boson emitted in the scattering of electrons against the nucleus. It sets a strong bound on the coupling to the electron for masses in the range $1.2 < \mphi < 52\MeV$. However, it too relies on the $\phi$ lifetime being in a certain range.  Ref.~\cite{Egli:1989vu} sets a very strong limit on the decay $\pi^+\rightarrow e^+\nu_e (\phi\rightarrow e^+e^-)$ and excludes masses in the range $10<\mphi < 110\MeV$, but again relies on the coupling to electrons. Finally, we note that any coupling to neutrinos must also be strongly suppressed as such an interaction will strongly affect the well measured interactions of neutrinos with matter. 

\section{Contributions to energy shifts in other muonic systems}

\begin{figure}
\begin{center}
\includegraphics[width=0.45 \textwidth]{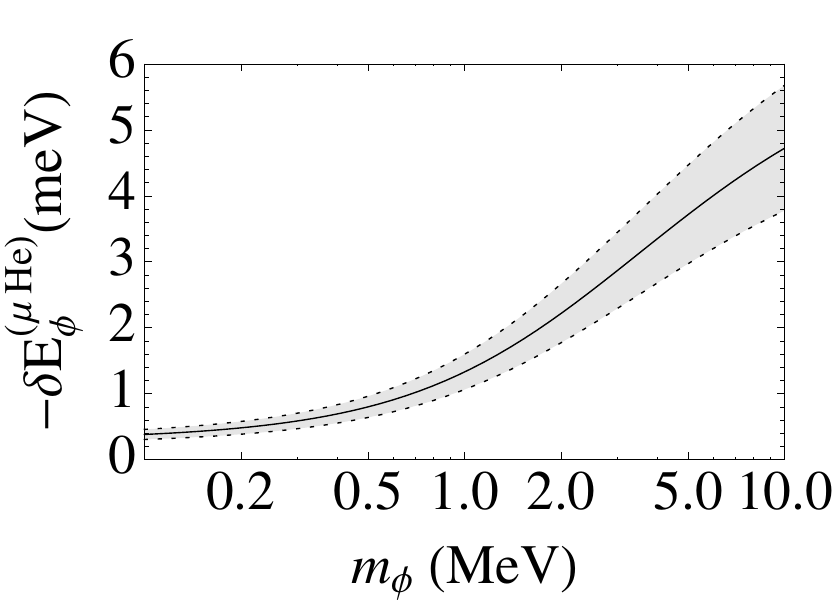}
\end{center}
\caption{The contribution to the  2S-2P energy splitting in muonic helium, Eq.~(\ref{eqn:HNenergyshift}), plotted against the mediator's mass. The curve is normalized to $\delta E_{\mu H} = -0.31\meV$, which is the discrepancy between the experimentally measured value and the theoretical prediction, assuming the CODATA value for the proton-radius, $r_p = 0.8768~\rm{fm}$. The dotted curves represent the $\pm1$ s.d. uncertainty about this value.}
\label{fig:deltaEHevsM}
\end{figure}

We now move on to discuss predictions concerning other muonic systems such as muonic deuterium and helium, and true muonium. Neglecting any possible coupling to the neutrons, the energy shift in the 2S-2P transition due to the new force is given by a simple generalization of Eq.~({\ref{eqn:energyshift}). Accounting for the change in the reduced mass and the atomic number, it can be written in terms of the contribution to the energy shift in muonic hydrogen,
\be
\delta E_\phi^{(\mu N)} = Z \frac{f(a_{\mu N}\mphi )}{f(a_{\mu H}\mphi )} \left( \frac{a_{\mu H}^3}{a_{\mu N}^3} \right)~\delta E_\phi^{(\mu H)}
\label{eqn:HNenergyshift}
\ee
where $N$ stands for the different possible nuclei (deuterium, helium, and etc.) and $Z$ is the atomic number. 

The most straightforward prediction is that, more or less independent of the mediator's mass, the muonic deuterium system, $\mu$-$D$, should exhibit almost exactly the same energy shift in the 2S-2P transition as the $\mu$-$p$ system. Depending on $\mphi$, it deviates from it by at most $\approx\pm15\%$ due to the change in the reduced mass, and  we therefore predict $\delta E_{\phi}^{(\mu D)} = -0.3\pm0.1\meV$. 

Next we consider the muonic helium system. The 2S$_{1/2}$-2P$_{3/2}$ and 2S$_{1/2}$-2P$_{1/2}$ level splittings in $(\mu^-~^4$He$)^+$ muonic ion were measured in the 70's and reported in refs.~\cite{Carboni:1976wb,Carboni:1977uw}. Unfortunately, these measurements were later criticized by other groups which cast doubt on the validity of the results. Experimentally~\cite{Hauser:1992zz}, the 2S$_{1/2}$-2P$_{3/2}$ resonance was not found at the frequency reported by ref.~\cite{Carboni:1976wb}. Theoretically~\cite{Landua:1982hp}, the lifetime of the 2S state is predicted to be too short at the pressure (40 bar) used in the experiment, a result which was later confirmed experimentally by ref.~\cite{Eckhause:1985cz}.  However, further measurements on muonic helium are planned in the near future at PSI~\cite{Pohl:2010zz}. In Fig.~\ref{fig:deltaEHevsM} we plot the prediction for the 2S-2P energy-shift in this system as a function of the mediator mass, $\mphi$.

\begin{figure}
\begin{center}
\includegraphics[width=0.45 \textwidth]{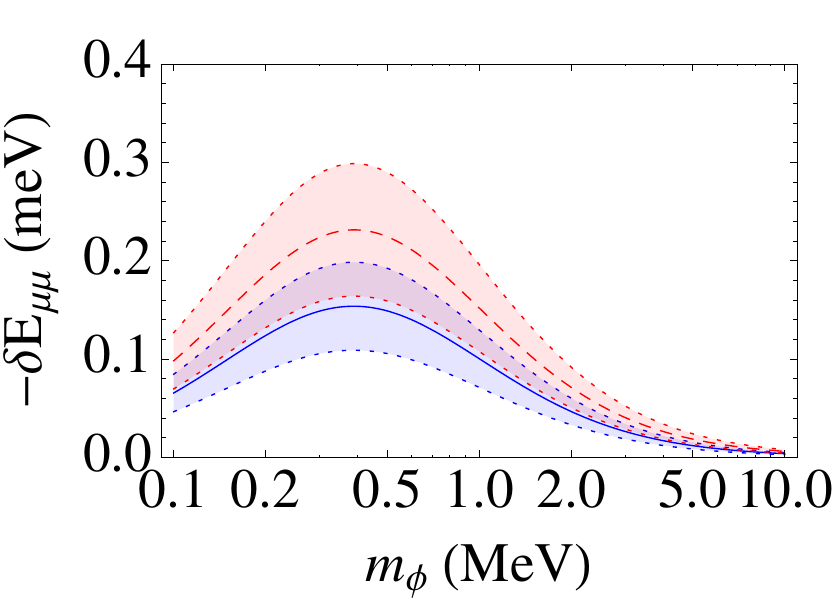}
\end{center}
\caption{The contribution to the energy shift in true muonium as a function of the force carrier mass. Curves are as in Fig.~\ref{fig:deltaEvsM}.}
\label{fig:deltaEmumuvsM}
\end{figure}


A new force that couples to muons would also contribute to the energy levels of true muonium ($\mu^+$-$\mu^-$ bound state), a system that is yet to be observed. In an ineresting recent proposal, the authors of Ref.~\cite{Brodsky:2009gx} have discussed two possible production mechanisms, $e^+e^-\rightarrow \mu^+\mu^-$ and $e^+e^-\rightarrow \gamma \mu^+\mu^-$, as particularly promising channels. If in addition to production, spectroscopic studies of true muonium can be achieved, then it may be possible to observe the small energy-shift due to the force carrier exchange. In Fig.~\ref{fig:deltaEmumuvsM} we plot the predicted energy shift in the  2S-2P transition when the muon coupling, $g_\mu$, is chosen so as to fix the discrepancy in $(g-2)_\mu$. This energy shift is much smaller than the expected level-splitting due to vacuum polarization and will undoubtedly be very difficult to observe. Nevertheless, it serves as an unambiguous prediction that may be tested in the future.

\section{Conclusions}

A new force carrier with a mass of $\approx\MeV$ that couples to both protons and muons with $g_{\mu,p}\approx 4\times 10^{-4}$ can explain the discrepancies observed in both $(g-2)_\mu$ and the muonic hydrogen 2S$_{1/2}$-2P$_{3/2}$ energy splitting. However, the coupling of such a new force to either neutrons or electrons is constrained by several past measurements. Nevertheless, none of these constraints exclude this possibility in a model-independent way. Assuming only couplings to muons and protons, this possibility lends itself to concrete predictions for the expected energy shifts in muonic deuterium and helium as well as true muonium. 
 
\textbf{Note added:} While this paper was being completed ref.~\cite{Barger:2010aj} appeared, which also considers new-physics contributions to  the Lamb shift in muonic hydrogen. 

\begin{acknowledgments}
We would like to thank N. Arkani-Hamed, A. Cohen, P. Schuster, N. Toro, J. Wacker, and N. Weiner for useful discussions. IY is supported by the James Arthur fellowship. DTS is supported by NSF grant PHY-0856522. IY would also like to thank the Perimeter Institute where this work was completed.   
\end{acknowledgments}
\bibliography{MeVForce}
\end{document}